\documentclass[fleqn, 12pt] {article}
\usepackage{epsf}
\begin{document}
\title{Combined effect of horizontal magnetic field and vorticity on  Rayleigh -Taylor instability }
\author{ Rahul Banerjee\thanks{e-mail:
rbanerjee.math@gmail.com} \\
 St Paul's Cathedral Mission College, 33/1, Raja Rammohan Roy
Sarani,\\ Kolkata – 700 009, India\\}
\date{}

\maketitle
\begin{abstract}
In this research, the height, curvature and velocity of the bubble
tip in Rayleigh-Taylor instability at arbitrary Atwood number with
horizontal magnetic field are investigated. To support the earlier
simulation and experimental results, the vorticity generation
inside the bubble is introduced. It is found that, in early
nonlinear stage, the temporal evolution of the bubble tip
parameters depend essentially on the strength and initial
perturbation of the magnetic field, although the asymptotic nature
coincides with the non magnetic case. The model proposed here
agrees with the previous linear, nonlinear and simulation
observations.
\end{abstract}
\newpage

\section*{I INTRODUCTION}
The Rayleigh-Taylor Instability occurs when a lighter density
fluid pushes the heavier one against the gravitational force
field. This instability appears in many physical and astrophysical
situations, such as Inertial Confinement Fusion, where the
magnetic field provides a stabilizing effect of the two fluid
instability${\cite{mrg10}}$, overturn of the outer portion of the
collapsed core of massive stars, etc. In the linear regime, the
perturbation grows exponentially with the growth rate
$\sqrt{Akg}$, where $A=\frac{\rho_h-\rho_l}{\rho_h+\rho_l}$ is the
Atwood number, $\rho_h$ and $\rho_l$ are the densities of heavier
and lighter fluid, respectively, $k$ is the perturbation wave
number and $g$ is the interfacial acceleration ${\cite{sc61}}$. In
the nonlinear stage, the interface can be divided into the bubble
of the lighter fluid rising into the heavier fluid, and spike of
the heavier fluid penetrating into the lighter fluid. There are
several methods for describing the nonlinear effect on this
instability. Among them, Layzer's ${\cite{dl55}}$ describes a
formulation where the interface near the tip of the bubble is
approximated by a parabola and determined the position, curvature
and velocity of the bubble tip. Extending this model, Goncharov
${\cite{vg02}}$ derived the asymptotic velocity of the bubble tip,
which is $\sqrt{\frac{2A}{1+A}\frac{g}{3k}}$. However, the
observed simulation and experimental results
${\cite{pr06,bt06,rb11}}$ indicate that nonlinear theory correctly
captures the bubble behavior in the early nonlinear phase, but
fails in the highly nonlinear stage. Betti and Sanz
${\cite{bt06}}$ shows that this occurs due to vorticity accretion
inside the bubble and the velocity of the bubble tip is slightly
higher than the classical value obtained by Goncharov
${\cite{vg02}}$.

In an Inertial Confinement Fusion situation or in the
astrophysical situation the fluid may be ionized or may get
ionized through laser irradiation in laboratory condition. In this
case the study of magnetic field effect on Rayleigh-Taylor
instability is needed. Under the linear theory, the influence of
magnetic field on Rayleigh-Taylor instability has been studied in
detail by Chandrasekhar ${\cite{sc61}}$. He observed that, when
the magnetic field is parallel to the interface separating of two
fluids, the growth rate of the Rayleigh-Taylor instability is
unaffected by magnetic field. However, using Layzer's model, Gupta
et. al${\cite{mrg10}}$ pointed that the parallel magnetic field
becomes a stabilizing factor of the instability.

The asymptotic growth, curvature and growth rate of the bubble tip
in Rayleigh-Taylor instability, which is one of the main factors
in Inertial Confinement Fusion or in laboratory experiments, have
been discussed by analytical and numerical approaches. In the
presence of magnetic field, the dynamics of the bubble tip has
been analyzed by considering the vorticity accumulation inside the
bubble. The magnetic field is assumed to be parallel to the plane
of the two fluid interface and acts in a direction perpendicular
to the wave vector. The basic model is based on the Layzer's
theory.

The structure of the paper is as follows. Section II describes the
kinematical and dynamical boundary conditions for the temporal
evolution of the bubble tip in Rayleigh-Taylor instability for
incompressible, inviscid fluids. Here the heavier fluid is assumed
to be irrotational where the lower one is rotational. The results
and discussions are presented in Section III.

\section*{II Basic Equations and Boundary Conditions}
We suppose that a fluid of density $\rho_h$ lies in the region
$z>0$ and that a second fluid of density $\rho_l$ lies in the
region $z<0$. The system is subject to a uniform acceleration $g$
in the negative direction of $z$ axis (see Fig.1). The magnetic
field is taken along the direction of $y$ axis, i.e, parallel to
the surface of separation.
\begin{eqnarray} \label{eq:1}
   \vec{B} = \left\{
     \begin{array}{lr}
      \widehat{y} B_h(x,z,t) & : z>0\\
       \widehat{y} B_l(x,z,t) & : z<0
     \end{array}
   \right.
\end{eqnarray}
According to the chosen magnetic field $\vec{\nabla}.\vec{B}=0$
everywhere.

Here we are considering two dimensional problem. Therefore we
approximate the perturbed interface by a parabola, given by
\begin{eqnarray}\label{eq:2}
z=\eta(x,t)=\eta_{0}(t)+\eta_{2}(t)x^2
\end{eqnarray}
where, for a bubble, $\eta_{0}(t)>0$ and $\eta_{2}(t)<0$.

The kinematical boundary conditions satisfied by the interfacial
surface $z=\eta(x,t)$ are
\begin{eqnarray}\label{eq:3}
\frac{\partial\eta}{\partial t}+v_{hx}
\frac{\partial\eta}{\partial x}=v_{hy}
\end{eqnarray}
\begin{eqnarray}\label{eq:4}
\frac{\partial\eta}{\partial x}(v_{hx}-v_{lx})=v_{hy}-v_{ly}
\end{eqnarray}
where $(v_{h,l})_{x,y}$ are the velocity components of the heavier
 and lighter  fluids, respectively.

The fluid motion is governed by the ideal magnetohydrodynamic
equations
\begin{eqnarray}\label{eq:5}
 \rho[\frac{\partial \vec{v}}{\partial
t}+(\vec{v}\cdot\vec{\nabla})\vec{v}]+\vec{\nabla}(p+g\rho
z)=\frac{1}{\mu}(\vec{\nabla}\times\vec{B})\times \vec{B}
=\frac{1}{\mu}(\vec{B}\cdot
\vec{\nabla})\vec{B}-\frac{1}{2\mu}\vec{\nabla}(B^2)=-\frac{1}{2\mu}\vec{\nabla}(B^2)
\end{eqnarray}
[$\frac{1}{\mu}(\vec{B}\cdot \vec{\nabla})\vec{B}=0$, as
$\vec{B}(x,z,t)$ is taken along the $y$ axis]
\begin{eqnarray}\label{eq:6}
\frac{\partial \vec{B}}{\partial t}=
\vec{\nabla}\times(\vec{v}\times \vec{B})
\end{eqnarray}
 According to Layzer's model ${\cite{dl55,rb12}}$, the velocity
potential describing the irrotational motion for the heavier fluid
is assumed to be given by
\begin{eqnarray}\label{eq:7}
\phi_h(x,z,t)=a(t) \cos(kx) e^{-k(z-\eta_{0}(t))}
\end{eqnarray}
with $\vec{v_{h}}=-\vec{\nabla}\phi_h$.

Since $\vec{\nabla}\cdot\vec{v_{h}}=0$, the equation of motion of
the upper incompressible fluid leads to the following integral
$\cite{rb11}$:
\begin{eqnarray}\label{eq:8}
\rho_{h}[-\frac{\partial \phi_{h}}{\partial t}+
\frac{1}{2}(\vec{\nabla} \phi_{h})^{2}+ g
z]+p_{h}+\frac{1}{2\mu_h}B_h^2=f_{h}(t)
\end{eqnarray}
For the lighter fluid the motion inside the bubble is assumed
rotational ${\cite{bt06}}$ with vorticity
$\vec{\omega}=(\frac{\partial v_{lz}}{\partial x}-\frac{\partial
v_{lx}}{\partial z}){\hat{y}}$. The motion is described by the
stream function $\Psi(x,y,t)$, given by
\begin{eqnarray}\label{eq:9}
\Psi(x,z,t)=b_{0}(t)x+[b_{1}(t)e^{k(z-\eta_{0})}+\omega_{0}(t)/k^2]
\sin{(k x)}
\end{eqnarray}
with $v_{lx}=-\frac{\partial \Psi}{\partial z}$ and
$v_{lz}=\frac{\partial \Psi}{\partial x}$.

Hence
\begin{eqnarray}\label{eq:10} \nabla^2 \Psi =-\omega
\end{eqnarray}
Let $\chi(x,z,t)$ be a function such that
\begin{eqnarray}\label{eq:11}
\nabla^2 \chi =-\omega
\end{eqnarray}
Hence $(\Psi-\chi)$ is a harmonic function as $\nabla^2
(\Psi-\chi)=0$. Let $\Phi(x,z,t)$ be its conjugate function
\begin{eqnarray}\nonumber
\frac{\partial \Phi}{\partial x}=\frac{\partial \Psi}{\partial
z}-\frac{\partial \chi}{\partial z}
\end{eqnarray}
\begin{eqnarray}\label{eq:12}
\frac{\partial \Phi}{\partial z}=-\frac{\partial \Psi}{\partial
x}+\frac{\partial \chi}{\partial x}
\end{eqnarray}
Thus the velocity components of the lighter fluid are
\begin{eqnarray}\nonumber
v_{lx}=-\frac{\partial \Psi}{\partial z}=-\frac{\partial
\Phi}{\partial x}-\frac{\partial \chi}{\partial z}
\end{eqnarray}
\begin{eqnarray}\label{eq:13}
v_{lz}=\frac{\partial \Psi}{\partial x}=-\frac{\partial
\Phi}{\partial z}+\frac{\partial \chi}{\partial x}
\end{eqnarray}
Using Eqs.(10)-(13), the first integral of the equation of motion
of the lighter fluid is given by
\begin{eqnarray}\label{eq:14}
\rho_{l}[-\frac{\partial \Phi}{\partial t}+
\frac{1}{2}(\vec{v}_{l} ){^2}-\omega \Psi +g z]+\int
\rho_{l}[(\Psi \frac{\partial \omega}{\partial z}-\frac{\partial
\dot{\chi}}{\partial z})dx +(\Psi \frac{\partial \omega}{\partial
x}+\frac{\partial \dot{\chi}}{\partial x})dz] +
p_{l}+\frac{1}{2\mu_l}B_l^2=f_{l}(t)
\end{eqnarray}
Here we set
\begin{eqnarray}\label{eq:15}
\chi(x,z,t)=\omega_{0}(t)\sin{(k x)}/k^2
\end{eqnarray}
Therefore Eq.(13) gives
\begin{eqnarray}\label{eq:16}
\Phi(x,z,t)=-b_0(t)y+b_1(t)\cos{(k x)} e^{k(z-\eta_0)}
\end{eqnarray}
From Eqs.(8) and (14), we obtain our dynamical boundary condition:
\begin{eqnarray}\label{eq:17}
\nonumber \rho_{h}[-\frac{\partial \phi_{h}}{\partial t}+
\frac{1}{2}(\vec{\nabla} \phi_{h})^{2}+ g z]-
\rho_{l}[-\frac{\partial \Phi}{\partial t}+
\frac{1}{2}(\vec{\nabla} \Phi){^2}-\omega \Psi +g z]-\int
\rho_{l}[(\Psi \frac{\partial \omega}{\partial z}-\frac{\partial
\dot{\chi}}{\partial z})dx \\+(\Psi \frac{\partial
\omega}{\partial x}+\frac{\partial \dot{\chi}}{\partial x})dz]
+(p_{h}-p_l)+(\frac{1}{2\mu_h}B_h^2-\frac{1}{2\mu_l}B_l^2)=f_{h}(t)-f_l(t)
\end{eqnarray}
satisfied at the interface $z=\eta(x,t)$.

Now we turn to our magnetic field equations. In virtue of Eqs. (1)
and (6) the magnetic fields are assumed to be
\begin{eqnarray}\label{eq:18}
B_h(x,z,t)=\beta_{h0}(t)+\beta_{h1}(t)e^{-k(z-\eta_{0})} \cos{(k
x)}
\end{eqnarray}
\begin{eqnarray}\label{eq:19}
B_l(x,z,t)=\beta_{l0}(t)+\beta_{l1}(t)e^{k(z-\eta_{0})} \cos{(k
x)}
\end{eqnarray}

Substituting $\eta(x,t)$, $v_{hx}$, $v_{hz}$, $v_{lx}$ and
$v_{lz}$ in Eqs. (3) and (4), and expanding in powers of the
transverse coordinate $x$  and neglecting terms $O(x^i)$ ($i\geq
3$), we obtain the following equations $\cite{rb12}$
\begin{eqnarray}\label{eq:20}
\frac{d\xi_{1}}{d\tau }=\xi_{3}
\end{eqnarray}
\begin{eqnarray}\label{eq:21}
\frac{d\xi_{2}}{d\tau}=-\frac{1}{2}(6\xi_{2}+1)\xi_{3}
\end{eqnarray}
\begin{eqnarray}\label{eq:22}
\frac{kb_{0}}{\sqrt{kg}}=\frac{6\xi_{2}(2\xi_{3}-\Omega)}{(6\xi_{2}-1)}
\end{eqnarray}
\begin{eqnarray}\label{eq:23}
\frac{k^2b_{1}}{\sqrt{kg}}=-\frac{(6\xi_{2}+1)\xi_{3}-\Omega}{(6\xi_{2}-1)}
\end{eqnarray}
where $ \xi_{1}=k\eta_{0}$, $\xi_{2}=\frac{\eta_{2}}{ k}$ and
$\xi_{3}=\frac{k^2a}{\sqrt{kg}}$ are the nondimensionalized bubble
height, curvature and velocity respectively, $\tau=t\sqrt{kg}$ is
the nondimensionalized time and
$\Omega=\frac{\omega_{0}}{\sqrt{kg}}$ is the nondimensionalized
vorticity.

Next substituting for the velocity components  $v_{hx}$, $v_{hz}$,
$v_{lx}$, $v_{lz}$ and $B_h(x,z,t)$, $B_l(x,z,t)$ in the  Eq.(6)
and equating coefficients of $x^i$ for $i=0$ and $2$ we obtain the
following four equations
\begin{eqnarray}\label{eq:24}
\dot{\beta}_{h0}+\dot{\beta}_{h1}=0\hskip5pt\mbox{i.e,
}\beta_{h0}+\beta_{h1}=B_{h0}\hskip5pt\mbox{(say)}
\end{eqnarray}
\begin{eqnarray}\label{eq:25}
\frac{d\xi_{4}}{d\tau}=\xi_3\xi_4\frac{2\xi_2-1}{2\xi_2+1}
\end{eqnarray}
and
\begin{eqnarray}\label{eq:26}
\dot{\beta}_{l0}+\dot{\beta}_{l1}=0\hskip5pt\mbox{i.e,
}\beta_{l0}+\beta_{l1}=B_{l0}\hskip5pt\mbox{(say)}
\end{eqnarray}
\begin{eqnarray}\label{eq:27}
\frac{d\xi_{5}}{d\tau}=\xi_5\frac{\xi_3(2\xi_2+1)(6\xi_2+1)+2\Omega(2\xi_2-1)}{(2\xi_2-1)(6\xi_2-1)}
\end{eqnarray}
where $\xi_4=\frac{\beta_{h1}}{B_{h0}}$ and
$\xi_5=\frac{\beta_{l1}}{B_{l0}}$.

Again the fluid pressures together with the magnetic pressures on
both sides of the interface are equal $\cite{mrg10}$, i.e.,
\begin{eqnarray}\label{eq:28}
p_h+\frac{B_{h0}^2}{2\mu_h}=p_l+\frac{B_{l0}^2}{2\mu_l}
\end{eqnarray}
Using Eq.(28) in Eq.(17), the coefficient of $x^2$ of Eq .(17)
gives the following equation for $\xi_3$.
\begin{eqnarray}\nonumber
\frac{d\xi_3}{d\tau}=\frac{1}{D(\xi_2,r)}[-N(\xi_2,r)\frac{\xi_3^2}{(6\xi_2-1)}+2(r-1)(6\xi_2-1)\xi_2
+\frac{\Omega^2-5(6\xi_{2}+1)\Omega\xi_3}{(1-6\xi_2)}+\dot{\Omega}]
\end{eqnarray}
\begin{eqnarray}\label{eq:27}
-\frac{6\xi_2-1}{D(\xi_2,r)}[rV_h^2\xi_4(2\xi_2+1)+V_l^2\xi_5(2\xi_2-1)]
\end{eqnarray}
where
\begin{eqnarray}\label{eq:28}
N(\xi_2,r)=36(1-r)\xi_{2}^{2}+12(4+r)\xi_{2}+(7-r)
\end{eqnarray}
\begin{eqnarray}\label{eq:29}
D(\xi_2,r)=12(1-r)\xi_{2}^{2}+4(1-r)\xi_{2}+(r+1)
\end{eqnarray}
$r=\frac{\rho_h}{\rho_l}$ and $V_{h(l)}=\frac{k B_{h0(l0)}^2
}{\rho_{h(l)}\mu_{h(l)}g}$ is the normalized Alfven velocity.

 Thus the magnetic field affected Rayleigh - Taylor instability induced
growth of the bubble tip is determined by the parameters
$\xi_1(t)$, $\xi_2(t)$, $\xi_3(t)$ as also the magnetic induction
perturbation $\xi_4(t)$ and $\xi_5(t)$ given by Eqs. (20), (21),
(29), (25) and (27).

\section*{III  Results and Discussions}
The system of equations given by Eqs. (20), (21), (29), (25) and
(27) for the fluid parameters show that the complete understanding
of the Rayleigh-Taylor instability is not possible without knowing
the dependence of the vorticity $\Omega(\tau)$ on $\tau$.
According to the simulation results obtained by Snaz and Betti
$\cite{bt06}$, we chose the $\Omega(\tau)$ in the following form
so that the time dependence of $\Omega(\tau)$ has approximate
qualitative agrement with the simulation results.
 \begin{eqnarray}\label{eq:30}
\Omega(\tau)=\frac{\Omega_{c}}{1+2\tanh(\tau_{0})}[\tanh(\tau_{0})(1+\tanh(\tau))+\tanh(\tau-\tau_{0})]
\end{eqnarray}
Clearly $\Omega(\tau)$ increases from $0$ and tends to an
asymptotic value $\Omega_c$ as $\tau \rightarrow\infty$. The
constants $\tau_0$ and $\Omega_c$ are adjusted accordingly to
Ref.$\cite{bt06}$. The plot for $\Omega(\tau)$ is shown in Fig.2.
It is clear from the figure that the $\tau_0=8$ and $\Omega_c=2$
give a good approximation of the simulation results.

To integrate the system of equations numerically, it is necessary
to know the initial value of the parameters. The initial interface
is assumed to be $z=\eta_0(t=0)\cos (kx)$. The expansion of the
interfacial function gives $(\xi_2)_{initial}$
$=-\frac{1}{2}(\xi_1)_{initial}$ where $(\xi_1)_{initial}$ is the
arbitrary perturbation amplitude. As the perturbation starts from
rest, we may consider $(\xi_3)_{initial}=0$. The initial value of
$(\xi_4)_{initial}$ and $(\xi_5)_{initial}$ depend upon the
initial magnetic induction perturbation.

To describe the steady flow in Rayleigh-Taylor instability, we
first consider $B_{h0}\neq 0$, $B_{l0}=0$. This situation may
happen when the heavier fluid is magnetized and the lighter is non
magnetic. In this case $\xi_5=0$ and $V_l=0$.  The numerical
results of  the bubble dynamics are presented in Fig.3. Fig.3
demonstrates that, in early nonlinear stage the growth ($\xi_1$),
curvature ($\xi_2$) and velocity ($\xi_3$) depend on the magnetic
field and initial magnetic induction perturbation. More precisely,
the growth of the bubble tip reduces for large $B_{h0}$ and
$\xi_4<0$. This observations are supported by blue ($V_h=1$,
$(\xi_4)_{initial}=-0.1$) and dot-dash ($V_h=1.5$,
$(\xi_4)_{initial}=-0.1$) lines in Fig.3. This  happens as the
instability driving pressure differences term
$2(r-1)(6\xi_2-1)\xi_2$ together with the vorticity term
$\frac{\Omega^2-5(6\xi_{2}+1)\Omega\xi_3}{(1-6\xi_2)}+
\dot{\Omega}$ is lowered or enhanced  by
$rV_h^2\xi_4(2\xi_2+1)(6\xi_2-1)$ according as $\xi_4 <$ or $>0$.
However, the asymptotic values of the growth rate and curvature
are unaffected by the magnetic field as $\xi_4 \rightarrow 0$ as
$\tau \rightarrow\infty$. The asymptotic values are given by
setting $\frac{d\xi_2}{d\tau}=0$ and $\frac{d\xi_3}{d\tau}=0$.

\begin{eqnarray}\label{eq:33}
\xi_2|_{asymptotic} = -\frac{1}{6}
\end{eqnarray}
\begin{eqnarray}\label{eq:34}
\xi_3|_{asymptotic}=\sqrt{\frac{2}{3}\frac{A}{1+A}+\frac{\Omega_{c}^2}{4}\frac{1-A}{1+A}}
\end{eqnarray}
Thus the asymptotic growth rate and curvature become the same as
in the nonmagnetic case$\cite{bt06,rb11}$. This result agrees the
nonlinear result obtained by Gupta et. al. $\cite{mrg10}$.

Next we consider the reverse situation of the above case, i.e
$B_{h0}= 0$, $B_{l0}\neq0$. This circumstance  may happen when the
heavier fluid is nonmagnetic and the lighter is ionized. It is
clear from the Eq.(29) that the instability driving pressure
difference term $2(r-1)(6\xi_2-1)\xi_2$ together with the
vorticity term
$\frac{\Omega^2-5(6\xi_{2}+1)\Omega\xi_3}{(1-6\xi_2)}+
\dot{\Omega}$ is now lowered or enhanced  by
$V_l^2\xi_5(2\xi_2-1)(6\xi_2-1)$ (note that
$-\frac{1}{6}\leq\xi_2<0$) according as $\xi_4
>$ or $<0$. This conclusion is supported by the Fig.4, where
the growth of the bubble tip reduces for large $V_l$ with
$(\xi_5)_{initial}>0$.  In asymptotic stage, $\xi_5\rightarrow 0$
as $\tau\rightarrow\infty$. This has the consequence that the
asymptotic growth rate of the bubble tip becomes the same as in
the nonmagnetic case.

Thus, in presence of horizontal magnetic field, which is
perpendicular to the plane of motion, the parameters of the bubble
tip such as growth, curvature and growth rate depend on the
strength of the magnetic filed and the initial magnetic
perturbation at the early nonlinear stage. However the asymptotic
values are coincided with the non magnetic case. Previously
$\cite{mrg10}$ nonlinear results show that the asymptotic growth
rate depends upon Alfven velocity of the lower fluid only by
considering irrotational motion in both fluids. However, due to
vorticity accretion inside the bubble, here we observed that the
asymptotic growth rate does not depend upon the Alfven velocity of
the both fluids.

\begin{figure}[p]
\vbox{\hskip 1.cm \epsfxsize=12cm \epsfbox{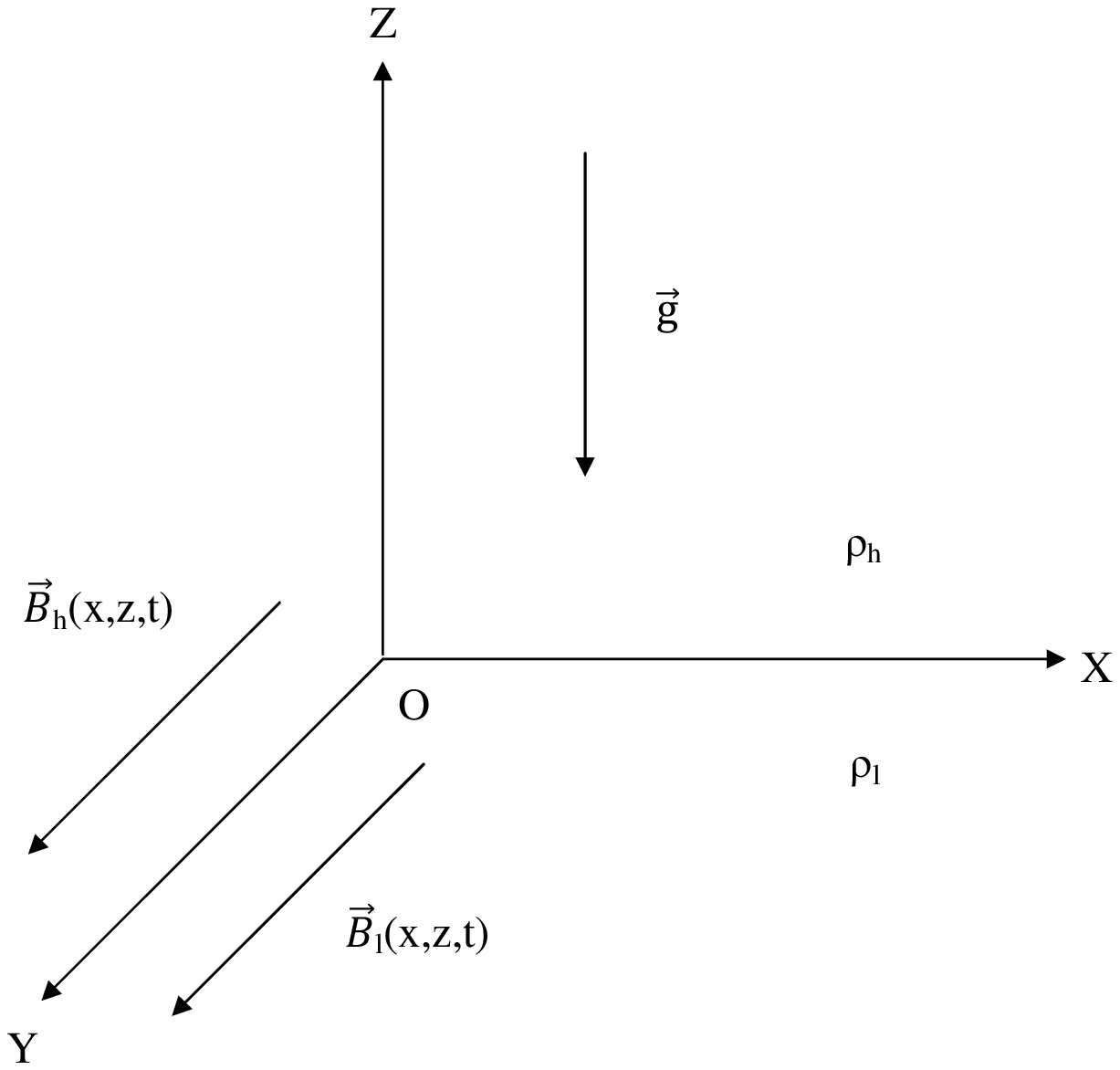}}
\begin{verse}
\vspace{-0.1cm} \caption{Schematic diagram of the model.
}\label{Fig:1}
\end{verse}
\end{figure}

\begin{figure}[p]
\vbox{\hskip 1.cm \epsfxsize=12cm \epsfbox{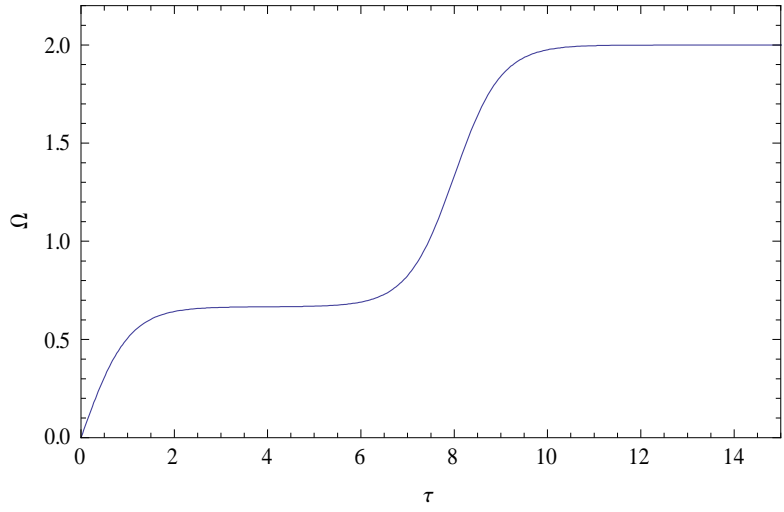}}
\begin{verse}
\vspace{-0.1cm} \caption{Vorticity $\Omega(\tau)$ plotted against
$\tau$ with asymptotic value $\Omega_{c}$=2 and parameter
$\tau_{0}$=8. }\label{Fig:2}
\end{verse}
\end{figure}

\begin{figure}[p]
\vbox{\hskip 1.cm \epsfxsize=12cm \epsfbox{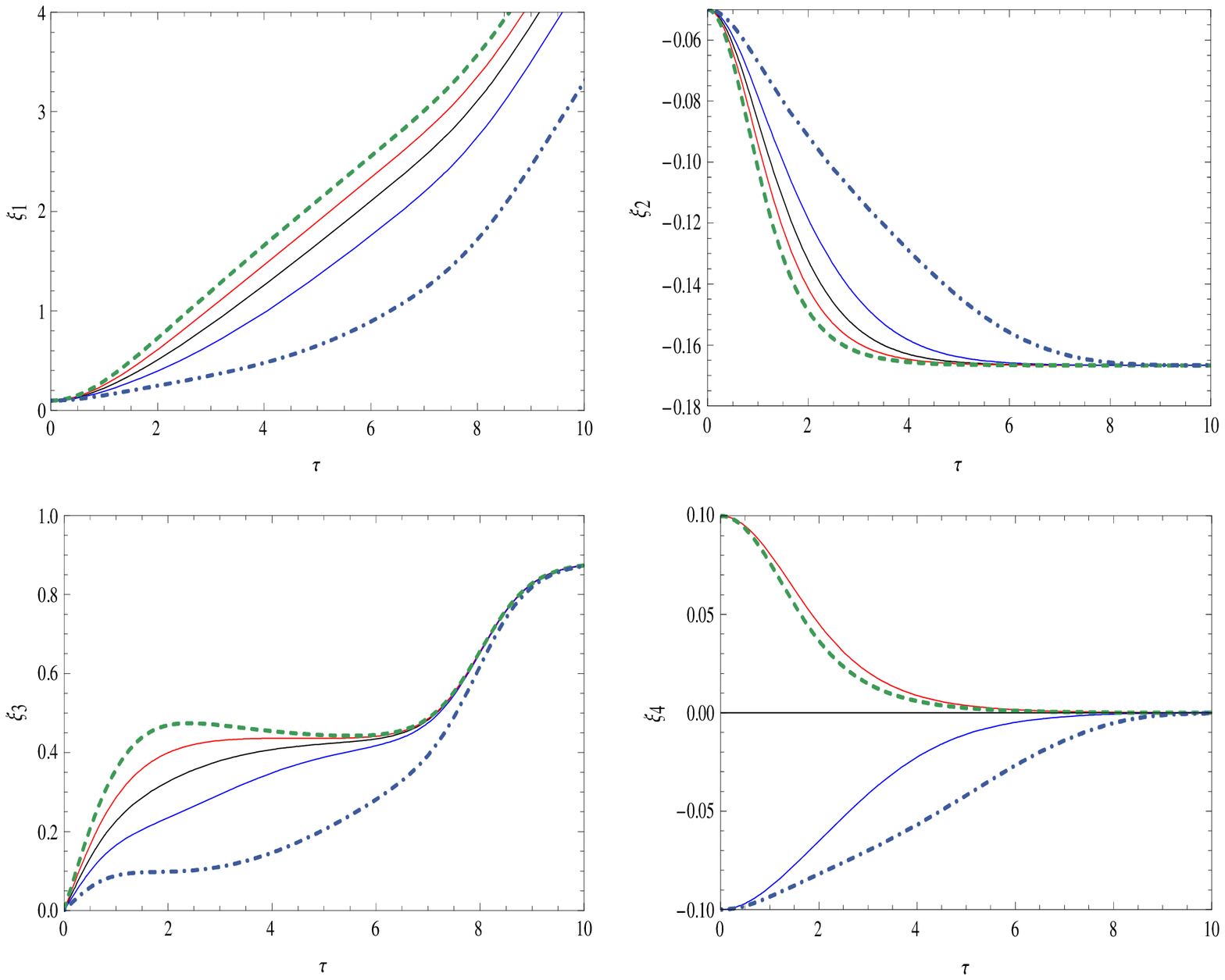}}
\begin{verse}
\vspace{-0.1cm} \caption{Variation of $\xi_1$, $\xi_2$, $\xi_3$,
$\xi_4$ with $\tau$ for $r=1.5$, $V_l=0$, $\Omega_{c}=2$,
$(\xi_1)_{initial}=0.1$, $(\xi_2)_{initial}=-0.05$ and
$(\xi_3)_{initial}=0$; ($V_h=0$, $(\xi_4)_{initial}=0$( Black
line), ($V_h=1$, $(\xi_4)_{initial}=0.1$( Red line),($V_h=1$,
$(\xi_4)_{initial}=-0.1$( Blue line),($V_h=1.5$,
$(\xi_4)_{initial}=0.1$( Dash line),($V_h=1.5$,
$(\xi_4)_{initial}=-0.1$( Dot-dash line).}\label{Fig:3}
\end{verse}
\end{figure}

\begin{figure}[p]
\vbox{\hskip 1.cm \epsfxsize=12cm \epsfbox{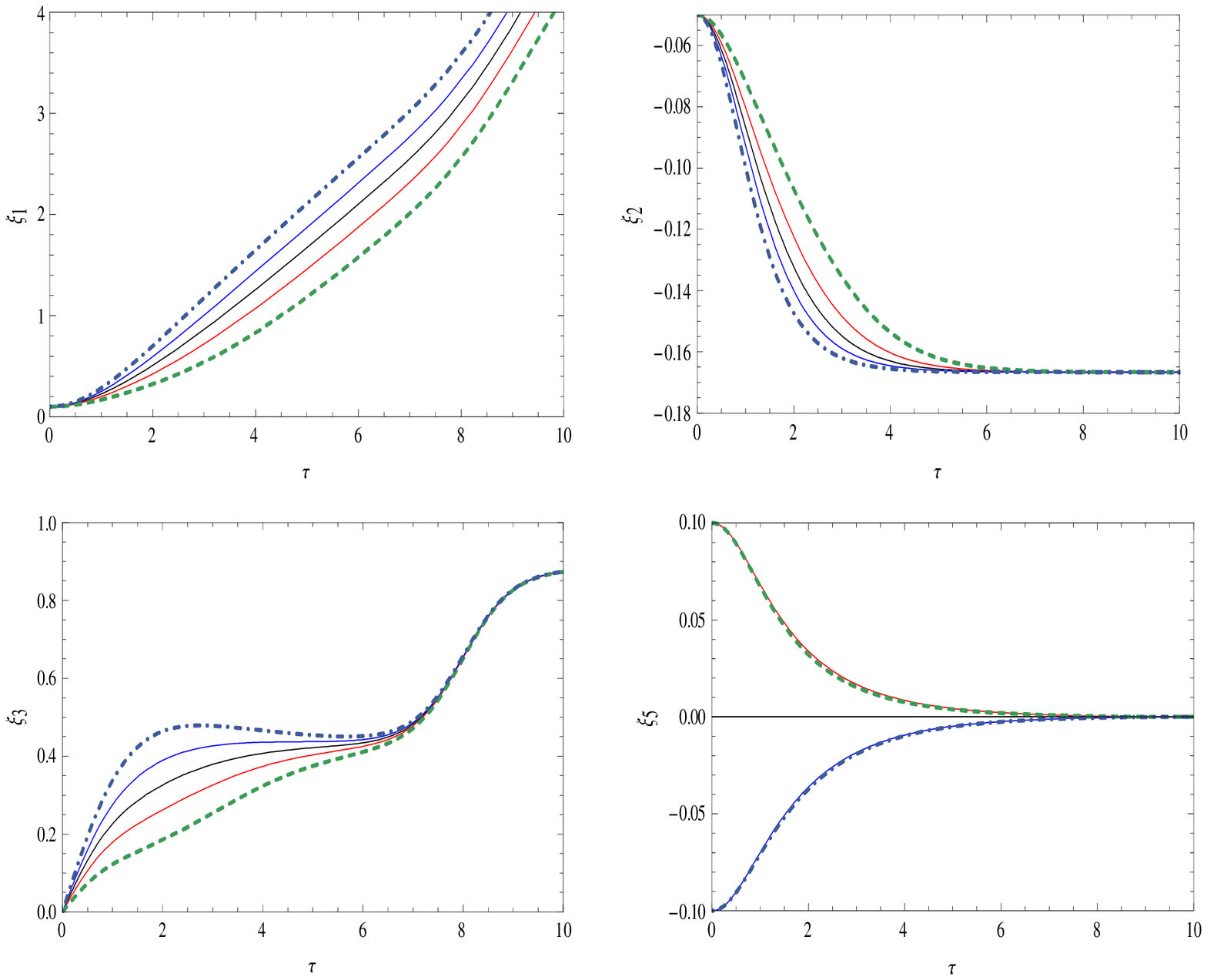}}
\begin{verse}
\vspace{-0.1cm} \caption{Variation of $\xi_1$, $\xi_2$, $\xi_3$,
$\xi_5$ with $\tau$ for $r=1.5$, $V_h=0$, $\Omega_{c}=2$,
$(\xi_1)_{initial}=0.1$, $(\xi_2)_{initial}=-0.05$ and
$(\xi_3)_{initial}=0$; ($V_h=l$, $(\xi_5)_{initial}=0$( Black
line), ($V_l=1$, $(\xi_5)_{initial}=0.1$( Red line),($V_l=1$,
$(\xi_5)_{initial}=-0.1$( Blue line),($V_l=1.5$,
$(\xi_5)_{initial}=0.1$( Dash line),($V_l=1.5$,
$(\xi_5)_{initial}=-0.1$( Dot-dash line).}\label{Fig:4}
\end{verse}
\end{figure}

\end{document}